\documentclass{article}
\usepackage{amsmath,amsthm}
\usepackage{amssymb,latexsym}
\usepackage[mathscr]{eucal}
\usepackage{setspace}
\usepackage{graphics}
\usepackage{array}
\usepackage[superscript,biblabel]{cite}

\begin{document}

\title{\textbf{Stability of Parker's Solar Wind Solution Near the Solar Surface}}
\author{Bhimsen K. Shivamoggi \\ University of Central Florida \\ Orlando, Fl 32816-1364}
\date{}
\maketitle

\begin{abstract}
 \normalsize Stability of Parker's \cite{Parker} steady solar wind solution near the solar surface is systematically investigated by posing a Sturm-Liouville problem for this solution. Parker's solar wind solution, whether it turns into a breeze or a supersonic wind depending on the pressure in the interstellar medium, is shown to describe the solar wind to start in a stable way as it leaves the solar surface from a state of,
\begin{itemize}
\item rest,
\item \textit{co-rotation} with the sun,
\item \textit{slow} motion
\end{itemize}
The isothermal gas flow assumption in Parker's solar wind model is then relaxed, and more realistic \textit{barotropic} fluid and \textit{diabatic} models are used for the gas flow. The stability of the solar wind flow, as it starts from a state of rest at the solar surface, is shown to continue to hold. Parker's solar wind solution therefore appears to be a stable attractor of this dynamical system.
\end{abstract}

\newpage

\section{Introduction}
Parker \cite{Parker} gave an ingenious stationary model which provided for the smooth acceleration of the stellar wind along open magnetic field lines through transonic speeds by continually converting the thermal energy into the kinetic energy of the wind. In the case of the Sun, the solar wind was confirmed and its properties were recorded by \textit{in situ} observations (see Meyer-Vernet \cite{Meyer}). The recent incoming data from the Parker Solar Probe has been providing a lot of useful information on the conditions in the solar corona (see Fisk and Kasper \cite{FiskKasper} and references thereof). The stellar rotation is found to lead to faster stellar winds and hence enable protostars and strong rotators to lose their angular momentum quickly via the mechanism of centrifugal and magnetic driving (Shivamoggi \cite{Shivamoggi}).

The stability of the steady solar wind solution given by Parker is an important question and some attempts have been made to shed light on the stability of Parker's supersonic solar wind solution (Jockers \cite{Jockers} ,Velli \cite{Velli6}). Here, we do a systematic investigation of stability of the Parker's solar wind solution by posing a Sturm-Liouville problem for the latter and show that the latter describes the wind to start in a stable way as it leaves the solar surface from a state of,
\begin{itemize}
\item rest,
\item \textit{co-rotation} with the sun,
\item \textit{slow} motion.
\end{itemize}
The isothermal gas flow assumption in Parker's \cite{Parker} solar wind model becomes weak upon including the extended active heating of the corona (Parker \cite{Parker7}). This is amended in a first approximation by using \textit{barotropic} fluid and \textit{diabatic} models for the gas flow and the solar wind flow is shown to start again in a stable way.

\section{Stability of the Parker Solar Wind Solution Near the Solar Surface}
Parker's hydrodynamic model \cite{Parker} assumes the solar wind to be represented by a spherically symmetric flow so the flow variables depend only on $r$, the distance from the star. The flow velocity $v$ is taken to be only in the radial direction - either inward (accretion model) or outward (wind model). We assume for analytical simplicity that the flow variables and their derivatives vary continuously so there are no shocks anywhere in the region under consideration. We now consider an unsteady version of Parker's model.

The mass conservation equation is

\begin{equation}
r^2 \frac{\partial \rho}{\partial t}+\frac{\partial}{\partial r} \left(r^2 \rho \right) v + r^2 \rho \frac{\partial v}{\partial r} = 0.
\label{eqn::1}
\end{equation}
where $\rho$ is the mass density.

Assuming the gravitational field to be produced by a central mass $M_s$, Euler's equation of momentum balance is,

\begin{equation}
\rho \left( \frac{\partial v}{\partial t}+v \frac{\partial v}{\partial r} \right) = -\frac{\partial p}{\partial r}- \frac{G M_s}{r^2}\rho
\label{eqn::2}
\end{equation}
$G$ being the gravitational constant, $M_s$ is the mass of the Sun, and $p$ is the pressure.

We assume the gas flow to occur under isothermal conditions, so

\begin{equation}
p = a^2 \rho
\label{eqn::3}
\end{equation}
$a$ being the constant speed of sound.

Near the solar surface, we assume the wind to start from a state of rest, we assume normal mode solutions of the form,

\begin{equation}
\begin{aligned}
\rho(r,t)=\rho_0(r)+ \hat{\rho}(r) e^{-i \omega t} \\
v(r,t) = \hat{v}(r) e^{-i \omega t}
\end{aligned}
\label{eqn::4}
\end{equation}
and assume the perturbations, denoted by hats overhead, to be small. Equations (\ref{eqn::1})-(\ref{eqn::3}), then give

\begin{equation}
\rho_o (r) = \chi e^{-\frac{G M_s}{a^2 r}}
\label{eqn::5}
\end{equation}
which is the Chapman \cite{Chapman} density profile, $\chi$ being a constant, and

\begin{equation}
\frac{d}{dr}\left(r^2 \frac{d \hat{\rho}}{dr}\right) + \frac{G M_s}{a^2} \frac{d \hat{\rho}}{dr}+ \frac{\omega^2}{a^2}(r^2 \hat{\rho})=0
\label{eqn::6}
\end{equation}
Equation (\ref{eqn::6}) may be rewritten as the \textit{Sturm-Liouville} equation,

\begin{equation}
\frac{d}{dr} \left[ f(r) \frac{d \hat{\rho}}{dr} \right] + \omega^2 g(r) \hat{\rho} = 0
\label{eqn::7}
\end{equation}
where,

\begin{equation}
\begin{aligned}
f(r) \equiv r^2 e^{-\frac{G M_s}{a^2 r}} > 0 \\
g(r) \equiv \frac{r^2}{a^2} e^{-\frac{G M_s}{a^2 r}}>0
\end{aligned}
\label{eqn::8}
\end{equation}

The \textit{Sturm-Liouville theory} (Birkhoff and Rota \cite{BirkhoffRota}) shows that the eigenvalues of a \textit{symmetric} operator (as in equation (\ref{eqn::7})) are \textit{real}. Therefore, $\omega^2$ is real.

Taking the complex conjugate of equation (\ref{eqn::7}), we have

\begin{equation}
\frac{d}{dr} \left[f(r) \frac{d \bar{\hat{\rho}}}{dr}\right] + \omega^2 g(r) \bar{\hat{\rho}}=0
\label{eqn::9}
\end{equation}

Assuming the boundary conditions,

\begin{equation}
r = r_s, \infty: \hspace{.1in} \hat{\rho}=0
\label{eqn::10}
\end{equation}
we obtain from equations (\ref{eqn::7}) and (\ref{eqn::9}),

\begin{equation}
- \int_{r_s}^{\infty}{f(r) \left| \frac{d \hat{\rho}}{dr}\right|^2}dr+\omega^2 \int_{r_s}^{\infty} {g(r) \left| \hat{\rho} \right|^2}dr = 0
\label{eqn::11}
\end{equation}
$r_s$ being the Sun's radius. Noting (\ref{eqn::8}), (\ref{eqn::11}) implies that $\omega^2 >0$, so $\omega$ is real. Thus, at the onset of instability, oscillatory motions prevail, and following Eddington \cite{Eddington} (see also Chandrasekhar \cite{Chand}), we have a case of \textit{overstability}. So, here the transition from stability to instability occurs via a marginal state exhibiting oscillatory motions. It may be mentioned that a similar situation prevails for Rayleigh-Taylor instability (Jukes \cite{Jukes}) and Kelvin-Helmholtz instability (Shivamoggi \cite{Shivamoggi14}) of a plasma in the presence of finite resistivity.

This establishes that Parker's solar wind solution describes the wind to actually start in a  stable way as it leaves the solar surface from a state of rest.

\section{Effect of Solar Rotation on Stability of the Parker Solar Wind Solution}
Assuming that the solar wind to be in a state of co-rotation with the Sun (the recent published observations by the Parker Solar Probe (Kasper et al. \cite{Kasper}) indicated solar wind to be coupled with solar rotation much more strongly than what was believed until now), Euler's equation of momentum balance now becomes (Shivamoggi \cite{Shivamoggi})

\begin{equation}
\rho \left( \frac{\partial v}{\partial t}+v \frac{\partial v}{\partial r} \right) = -a^2 \frac{\partial \rho}{\partial r} - \rho \frac{G M_s}{r^2}+ \rho \Omega_*^2 r
\label{eqn::12}
\end{equation}
$\Omega_*$ being the angular velocity of the Sun.

Near the solar surface, assuming again solutions to the perturbed state of the form given by (\ref{eqn::4}), and assuming the perturbations to be small, we obtain in place of equation (\ref{eqn::7}),

\begin{equation}
\frac{d}{dr} \left[ \tilde{f}(r) \frac{d \hat{\rho}}{dr} \right] + (\omega^2-3 \Omega_*^2) \tilde{g}(r) \hat{\rho} = 0
\label{eqn::13}
\end{equation}
where,

\begin{equation}
\begin{aligned}
\tilde{f}(r) \equiv r^2 e^{-\frac{1}{a^2}\left( \frac{G M_s}{r}+\frac{\Omega_*^2 r^2}{2}\right)}>0 \\
\tilde{g}(r) \equiv \frac{r^2}{a^2} e^{-\frac{1}{a^2}\left(\frac{G M_s}{r}+ \frac{\Omega_*^2 r^2}{2} \right)}>0
\end{aligned}
\label{eqn:: 14}
\end{equation}

The Sturm-Liouville theory (Birkhoff and Rota \cite{BirkhoffRota}) again implies $(\omega^2 - 3 \Omega_*^2)$ is real.

Assuming the boundary conditions (\ref{eqn::10}), we may again obtain from equation (\ref{eqn::13}),

\begin{equation}
- \int_{r_s}^{\infty} {\tilde{f}(r) \left| \frac{d \hat{\rho}}{dr} \right|^2} dr+(\omega^2 - 3 \Omega_*^2) \int_{r_s}^{\infty} {\tilde{g}(r) \left| \hat{\rho} \right|^2}dr = 0.
\label{eqn::15}
\end{equation}
Noting (\ref{eqn:: 14}), (\ref{eqn::15}) implies that, $\omega^2 > 0$, if $(\omega^2 - 3 \Omega_*^2)>0$, so $\omega$ is real. On the other hand, if $(\omega^2-3 \Omega_*^2)<0$, then (\ref{eqn::15}) implies $\hat{\rho} \equiv 0$. So, the perturbations remain bounded in either case, and Parker's solar wind solution is again shown to describe the wind to start in a stable way as it leaves the solar surface from a state of rest.

\section{Effect of Slow Initial Motion on Stability of the Parker Solar Wind Solution}

Assuming that the solar wind leaves the solar surface from a state of \textit{slow} motion, we assume solutions of the form,

\begin{equation}
\begin{aligned}
\rho(r,t) = \rho_0(r)+\hat{\rho}(r) e^{i \omega t} \\
v(r,t) = v_0(r)+\hat{v}(r) e^{i \omega t}
\end{aligned}
\label{eqn::16}
\end{equation}

Assuming $v_0 \ll a$ and $\sqrt{G M_s/r}$, equations (\ref{eqn::1}) and (\ref{eqn::3}) then give for the \textit{equilibrium} state, to first approximation,
\begin{equation}
\frac{\partial}{\partial r} (r^2 \rho_0 v_0) = 0 \hspace{.1in} or \hspace{.1in} r^2 \rho_0 v_0 = const = \sigma
\label{eqn::17}
\end{equation}
\begin{equation}
a^2 \frac{\partial \rho_0}{\partial r}-\rho_0 \frac{G M_s}{r^2} \approx 0
\label{eqn::18}
\end{equation}
Equations (\ref{eqn::17}) and (\ref{eqn::18}) give,

\begin{equation}
\rho_0 \approx \chi e^{-\frac{G M_s}{a^2 r}}, \hspace{.1in} v_0 \approx \frac{\sigma}{\chi r^2}e^{\frac{G M_s}{a^2 r}}
\label{eqn::19}
\end{equation}

Using (\ref{eqn::19}), we then obtain in place of equation (\ref{eqn::6}),
\begin{equation}
\frac{d}{dr} \left[ \xi(r) \frac{d \hat{\rho}}{dr} \right]+ \omega^2(1+i \mu) \eta(r) \hat{\rho}=0
\label{eqn::20}
\end{equation}
where,

\begin{equation}
\begin{aligned}
\xi(r) \equiv r^2 e^{-\frac{G M_s}{a^2 r}-\frac{i \omega}{G M_s} r^2 v_0}, \hspace{.1in} \mu(r) \equiv 2 \frac{v_0}{r \omega}>0\\
\eta(r) \equiv \frac{r^2}{a^2} e^{-\frac{G M_s}{a^2 r}-\frac{i \omega}{G M_s} r^2 v_0}
\end{aligned}
\label{eqn::21}
\end{equation}

Assuming the boundary conditions (\ref{eqn::10}), we may again obtain from equation (\ref{eqn::20}),

\begin{equation}
- \int_{r_s}^\infty (\xi + \bar{\xi}) |\frac{d\hat{\rho}}{dr}|^2 dr+\omega^2 \int_{r_s}^\infty \left[ (\eta+\bar{\eta})+ i \mu (\eta - \bar{\eta}) \right] |\hat{\rho}|^2 dr = 0.
\label{eqn::22}
\end{equation}
Noting from (\ref{eqn::21}) that $(\xi+\bar{\xi})$ and $(\eta+\bar{\eta})$ are real and greater than zero, and $(\eta - \bar{\eta})$ is imaginary with $(\eta-\bar{\eta})=-i \kappa$, $\kappa>0$, equation (\ref{eqn::22}) shows that $\omega^2$ is real and greater than zero. So, $\omega$ is real, and hence the perturbations remain bounded, and Parker's solar wind solution is again shown to describe the wind to start in a stable way as it leaves the solar surface from a state of \textit{slow} motion.

\section{Barotropic Fluid Model for the Parker Solar Wind}
A main assumption in Parker's  \cite{Parker} solar wind model that the gas glow occurs under isothermal conditions, expressed by (\ref{eqn::3}), becomes weak upon including the extended active heating of the corona (Parker \cite{Parker7}). This may be amended in a  first approximation by using  a barotropic \cite{Note1} fluid model for the gas flow. The barotropic model assumes that $p$ is a \textit{single-valued} function of $\rho$, so that
\begin{equation}
\frac{1}{\rho} \nabla p = \nabla \mathord{h} = a^2 \nabla \rho
\label{eqn::23}
\end{equation}
where $\mathord{h}$ represents the \textit{specific enthalpy}.

Near the solar surface, assuming again, solutions to the perturbed state of the form given by (\ref{eqn::4}), and assuming the perturbations to be small, we obtain in place of equation (\ref{eqn::7}),
\begin{equation}
\frac{d}{dr} \left[\hat{f}(r) \frac{d \hat{\rho}}{dr} \right]+ \omega^2 \hat{g}(r) \hat{\rho} = 0
\label{eqn::24}
\end{equation}
where,
\begin{equation}
\begin{aligned}
\hat{f}(r) = r^2 a^2 e^{G M_s \int \frac{dr}{r^2 a^2}}>0 \\
\hat{g}(r) = r^2 e^{G M_s \int \frac{dr}{r^2 a^2}}>0.
\end{aligned}
\label{eqn::25}
\end{equation}

The Sturm-Liouville theory (Birkhoff and Rota \cite{BirkhoffRota}) again implies that $\omega^2$ is real.

Assuming the boundary conditions (\ref{eqn::10}), we may again obtain from equation (\ref{eqn::24}),

\begin{equation}
- \int_{r_s}^\infty \hat{f}(r) \left| \frac{d \hat{\rho}}{d r} \right|^2 dr+ \omega^2 \int_{r_s}^\infty \hat{g}(r) \left| \hat{\rho} \right|^2 dr = 0
\label{eqn::26}
\end{equation}
Noting (\ref{eqn::25}), (\ref{eqn::26}) implies that $\omega^2 >0$, so $\omega$ is real. This establishes that Parker's solar wind flow starts again from the solar surface in a stable way also for the barotropic case. For the isothermal case, (\ref{eqn::24}) - (\ref{eqn::26}) reduce to (\ref{eqn::8}), (\ref{eqn::9}), and (\ref{eqn::11}).

\section{Diabatic Flow in the Parker Solar Wind}
Consider a diabatic flow in Parker's \cite{Parker} solar wind model, so the contribution from the extended heating of the corona (Parker \cite{Parker7}) may now be explicitly included. Such a diabatic heat input is required to sustain acceleration of the solar wind in a non-isothermal situation (Cranmer and Winebarger \cite{Cranmer}).

We have from the First Law of Thermodynamics, on assuming that the solar wind passes through a succession of equilibrium states,
\begin{equation}
dQ = d \mathord{h} - \frac{1}{\rho}dp-\frac{G M_s}{r^2}
\label{eqn::27}
\end{equation}
where $Q$ is the heat transferred to the wind. Assuming a perfect gas, with
\begin{equation}
p = \rho R T
\label{eqn::28}
\end{equation}
$T$ being the temperature and $R$ being the gas constant, we have
\begin{equation}
\mathord{h} = c_p T
\label{eqn::29}
\end{equation}
where $c_p = \frac{\gamma R}{\gamma -1}$ is the specific heat at constant pressure and $\gamma$ is the ratio of specific heats, $\gamma = c_p/c_v$, $c_v$ being the specific heat at constant volume.

Using (\ref{eqn::28}), and (\ref{eqn::29}), (\ref{eqn::27}) gives
\begin{equation}
\frac{\partial T}{\partial r} = \frac{1}{c_p}\left( \frac{\partial Q}{\partial r}+ \frac{1}{\rho} \frac{\partial p}{\partial r}+ \frac{G M_s}{r^2} \right)
\label{eqn::30}
\end{equation}
Using equation (\ref{eqn::30}), we obtain from (\ref{eqn::28}),
\begin{equation}
\frac{\partial p}{\partial r} = a^2 \frac{\partial \rho}{\partial r} - (\gamma-1) \rho \frac{\partial Q}{\partial r}-(\gamma-1)\rho \frac{G M_s}{r^2}
\label{eqn::31}
\end{equation}
where,
$$
a^2 \equiv \frac{\gamma p}{\rho}
$$

Using equation (\ref{eqn::31}), equation (\ref{eqn::2}) leads to 
\begin{equation}
\rho \left( \frac{\partial v}{\partial t}+v \frac{\partial v}{\partial r} \right) = -a^2 \frac{\partial \rho}{\partial r}- \left[ (\gamma-1) \frac{\partial Q}{\partial r} + \gamma \frac{G M_s}{r^2} \right] \rho.
\label{eqn::32}
\end{equation}

Near the solar surface, assuming again, solutions to the perturbed state of the form given by (\ref{eqn::4}), and assuming the perturbations to be small, we obtain in place of equation (\ref{eqn::7}),
\begin{equation}
\frac{d}{dr} \left[ \tilde{f}(r) \frac{d \hat{\rho}}{dr} \right]+ \omega^2 \tilde{g}(r) \hat{\rho} = 0
\label{eqn::33}
\end{equation}
where,
\begin{equation}
\begin{aligned}
\tilde{f}(r) \equiv a^2 r^2 e^{\int \frac{1}{a^2 r^2} \left[ \gamma G M_s + (\gamma -1) \frac{\partial Q}{\partial r} r^2 \right]dr} >0 \\
\tilde{g}(r) \equiv \left[ \omega^2 r^2+2(\gamma-1) \frac{\partial Q}{\partial r} r \right]e^{\int \frac{1}{a^2 r^2}\left[ \gamma G M_s+(\gamma -1) \frac{\partial Q}{\partial r}r^2 \right] dr}>0
\end{aligned}
\label{eqn::34}
\end{equation}
noting $\partial Q/\partial r >0$, thanks to coronal heating.

The Sturm-Liouville theory (Birkhoff and Rota \cite{BirkhoffRota}) again implies that $\omega^2$ is real.

Assuming the boundary conditions (\ref{eqn::10}), we may again obtain from equation (\ref{eqn::33}),
\begin{equation}
- \int_{r_s}^\infty \tilde{f}(r) \left| \frac{d \hat{\rho}}{d r} \right|^2 dr+ \omega^2 \int_{r_s}^\infty \tilde{g}(r) \left| \hat{\rho} \right|^2 dr = 0
\label{eqn::35}
\end{equation}
Noting (\ref{eqn::34}), (\ref{eqn::35}) implies that $\omega^2>0$, so $\omega$ is real. This establishes that Parker's solar wind flow starts again from the Solar surface in a stable way also for a diabatic case.

\section{Discussion}
In this paper, we have done a systematic investigation of stability of Parker's steady solar wind solution near the solar surface by posing a Sturm-Liouville problem for this solution. We have found that Parker's solar wind solution, whether it turns into a breeze or a supersonic wind depending on the pressure in the interstellar medium, describes a stable start for the wind as it leaves the solar surface from a state of
\begin{itemize}
\item rest,
\item \textit{co-rotation} with the Sun,
\item \textit{slow} motion.
\end{itemize}
The isothermal gas flow assumption in Parker's \cite{Parker} solar wind model is then relaxed, and a more realistic barotropic fluid and diabatic models are used for the gas flow. Parker's \cite{Parker} solar wind flow is shown to start again from the solar surface in a stable way. Parker's solar wind solution therefore appears to be  a stable attractor of this dynamical system (Cranmer and Winebarger \cite{Cranmer}) and all other possible solar wind solutions are most likely unstable (Velli \cite{Velli9}). This undercuts previous conjectures alluding to a dynamical instability of Parker's solar wind solution (Parker \cite{Parker18}) driven by a spurious speculation that the latter is "finely tuned" in assuming the solar wind to accelerate smoothly through the sonic conditions exactly at the critical point $r_* \equiv G M_s/2 a^2$.

\section*{Acknowledgments}
I am thankful to Professor Eugene Parker for his helpful advice and suggestions on the stellar wind problem. My thanks are due to Professor Marco Velli for drawing my attention to the issue of stability of Parker's solar wind solution.

\end{document}